# IoT Architectures for Indoor Radon Management: a Prospective Analysis

Oscar Blanco-Novoa[1,2], Paulo Barros[3], Paula Fraga-Lamas[1,2], Sérgio Ivan Lopes[3,4,5], and Tiago M. Fernández-Caramés[1,2]

[1] Department of Computer Engineering, Faculty of Computer Science, Universidade da Coruña, 15071 A Coruña, Spain
[2] Centro de Investigación CITIC, Universidade da Coruña, 15071 A Coruña, Spain
o.blanco@udc.es, paula.fraga@udc.es, tiago.fernandez@udc.es
[3] ADiT-Lab, Instituto Politécnico de Viana do Castelo, 4900-348 Viana do Castelo, Portugal
paulobs@ipvc.pt, sil@estg.ipvc.pt
[4] CiTin - Centro de Interface Tecnológico Industrial, 4970-786 Arcos de Valdevez, Portugal
[5] IT - Instituto de Telecomunicações, Campus Universitário de Santiago, 3810-193 Aveiro, Portugal

**Abstract.** The demand for real-time Indoor Air Quality (IAQ) management has increased recently, since low-cost and modern sensors such as Particulate Matter (PM), Volatile Organic Compounds (VOCs), Carbon Monoxide (CO), Carbon Dioxide ($CO_2$), Radon (Rn), among others, have been put forward with considerable accuracy. Although these low-cost sensors cannot be considered measurement instruments, they are very useful for a vast number of application domains, such as home automation, smart building management, IAQ management, risk exposure assessment, to name a few. This paper presents a literature review and a prospective analysis and discussion regarding Internet of Things (IoT) technologies adopted to deal with scenarios that present known indoor Radon gas problems. Specifically, the main requirements for developing IoT-enabled radon management solutions are reviewed. Thus, a traditional IoT architecture is described, its main components are analyzed and some of the most recent academic solutions are reviewed. Finally, novel approaches for deploying IoT radon management architectures are presented together with the most relevant open challenges. In this way, this article presents a holistic review of the past, present, and future of indoor radon management in order to provide guidelines for future designers and developers.

## 1 Introduction

As humans spend more time indoors, it becomes more important to monitor Indoor Air Quality (IAQ) and manage the indoor environment. Clinical pathologies such as asthma and chronic obstructive pulmonary disease (COPD) are two respiratory conditions that have recently emerged as widespread illnesses with a known relation to poor IAQ performances. In addition, improving IAQ  has proven to be an effective measure for reducing the aerosol transmission of COVID-19.

The use of low-cost sensor-based Internet of Things (IoT) technologies not only promotes real-time and distributed indoor pollution monitoring, but also fosters novel paradigms for managing IAQ on indoor environments. This allows building managers to mitigate exposure  risks.

Low-cost and modern IAQ commercial sensor systems show high availability and an accuracy that may not make them measurement instruments, but which make them useful for a vast number of application domains [1], such as home automation, smart building management or risk exposure assessment, among others.

Radon gas is a radioactive element that cannot be detected by the human senses. It has been proven that it is a carcinogen and is a common air pollutant inside buildings that can be reduced via IAQ solutions.

As it is indicated at the bottom of Fig. 1, multiple radon gas sources can exist in a building, although such a gas usually comes from the soil through cracks and ventilation areas in the building or through its piping. Prospective radon management solutions have been gradually making their way, especially in the last decade, where many innovative scientific works related to indoor radon monitoring and mitigation have been presented. On an economic and political level, the European Union (EU) has also carried out its work with the publication of the Council Directive 2013/59/EURATOM. As a consequence, most EU member states have transposed such a directive to national regulations and have defined radon action plans to support policy decisions in this matter [2].

This article presents a prospective analysis of IoT architectures that can be adopted to deal with indoor scenarios with known indoor radon issues. Specifically, Section 2 analyzes the state of the art on the latest IoT technologies for indoor radon management. Section 3 describes the traditional IoT architectures for indoor radon management. Section 4 describes the latest architectures and points out the most relevant future directions for indoor radon management. Finally, Section 5 is dedicated to the conclusions.

## 2   Related Works

This section reviews  the state-of-the-art and related work on IoT technologies for indoor radon management. The performed analysis  is  focused  on articles and publications with integrated radon monitoring and mitigation systems that include active detection, data communications networks, cloud-based architectures, along with clear and concise descriptions of the techniques and technologies that were used.

In November 2021, Barros et al. [2] presented a literature review that compiles and compares the most relevant features of recent IoT technologies for indoor radon gas exposure management (i.e., for monitoring, assessment, and mitigation). This review also summarizes the main challenges and opportunities in the topic. All selected works included a cloud-based computing approach that follows the general architecture of IoT systems detailed in ITU-T Recommendation Y.4113 [3], which is composed of 3 blocks:

- **IoT Area Network**: It includes all sensors, actuators, and microcontroller units (MCUs) with storage and bidirectional communication capabilities.
- **Access network**: It includes the communication systems that allow IoT devices and gateways to connect with core networks through different technologies like optical fiber or radio access technologies (e.g., Wi-Fi, LoRa, BLE, ZigBee, 4G/LTE).
- **Core network**: It includes the physical infrastructure that is responsible for interconnecting the access network with service providers. It provides connectivity to web servers (for data processing and local or online storage), it allows for the visualization of the received data, it analyzes the impact of pollutants on target facilities, and it indicates end users radon gas levels and IAQ in general, among other information.

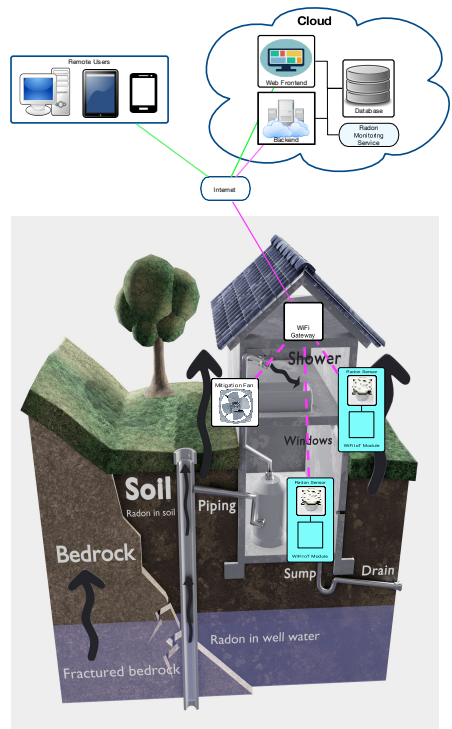

Fig. 1: Traditional cloud communications architecture for radon gas monitoring.

Table 1 identifies six different types of radon sensors presented by Barros et al. in [2], which use different radon gas detection techniques. Three of them are commercial sensors, while the other three are academic prototypes.

Table 1: Radon gas detection techniques and sensors.

| Radon detection technique | Radon sensor | Reference |
|---|---|---|
| Current or pulse ionization chambers | Radon FTLab RD200M | [4–10] |
| | Radon sensor from MidDec Scandinavia AB | [11] |
| Photodiode detection of alpha particles | Safety Siren Pro Series 3 | [12] |
| | Algade ÆR Plus | [13] |
| | Teviso RN53 (PIN diode) | [14] |
| | SARAD Radon Scout sensor | [15] |

The preferred architectures for connecting sensors include different MCUs and interfaces, being one of the most popular the ones from Espressif, Arduino and Raspberry Pi solutions, as it can be observed in Table 2.

The adopted communication technologies for real-time data collection are shown in Table 3. The most popular are SigFox, LoRaWAN and Wi-Fi, being the latter widely preferred.

Table 2: Hardware development kit.

| Brand | IoT DevKit | Reference |
|---|---|---|
| Espressif Systems | ESP8266 | [4–7] |
| | WeMos Mini D1 ESP8266 Wi-Fi ESP-12F | [12] |
| Arduino | Arduino MKR | [8, 9] |
| | Arduino UNO Rev3 | [10] |
| Raspberry Pi | Raspberry Pi 3 A + | [11] |
| | Raspberry Pi 3 Model B v1.2 | [15] |

Table 3: Communication technologies.

| Wireless Protocols (available) | Wireless Communications (used) | Reference |
|---|---|---|
| Wi-Fi | Wi-Fi (license-free 2.4 GHz and 5 GHz) | [10–12, 14, 15] |
| Wi-Fi and LoRaWAN | LoRaWAN (licence-free sub-1 GHz) | [4–7] |
| Sigfox and Wi-Fi | Sigfox (licence-free sub-1 GHz: 868 MHz) | [8, 9] |
| LoRaWAN | LoRaWAN (licence-free sub-1 GHz) | [13] |

Table 4: Summary of the most relevant features.

| Feature | Reference |
| --- | --- |
| Monitors radon concentration | [4–15] |
| Monitors hygrometric variables | [4–11, 13, 14] |
| Monitors other IAQ pollutants ($CO_2$, TVOC, etc.) | [4–9] |
| Sends alerts based on a pre-defined threshold | [4–10, 12, 13] |
| Includes risk prediction model and/or radon risk predictive algorithms | [9] |
| Includes a holistic risk prediction model and/or predictive algorithms (considering thermal comfort, energy efficiency, etc.) | none |
| Includes risk mitigation through human intervention | [4–8, 11–15] |
| Includes active risk mitigation through actuator devices | [9–11, 14] |
| Includes autonomous ventilation system | [9, 10] |

Table 4 summarizes the most important characteristics of the analyzed systems. Most studies follow a non-mechanical approach to IAQ monitoring. Mitigation actions are mostly taken by human operators after receiving alert messages, except for the study carried out by Alvarellos et al. [9], which already considers active mitigation and predictive risk prevention models.

These works show that, currently, IoT technologies are essential to improve IAQ, and can contribute to the so-called cognitive or intelligent buildings, where human intervention is no longer necessary, being replaced by autonomous building management systems. This paves the way for a holistic vision of the future, boosting critical factors, such as IAQ, energy efficiency and thermal comfort, which must be aligned with the use of indoor spaces (occupied rooms versus empty rooms). By considering these factors together, it is possible to improve the health and quality of life of buildings' residents.

The main challenges and opportunities of the previously mentioned technologies are related to the way they operate, the type of detection mechanisms they use, the type of system architecture, and the auxiliary communication components and technologies. Practical implementations require careful planning and an extensive list of requirements. These implementations should perform a meticulous selection of development kits that guarantee accuracy (for MCUs, sensors, and actuators), and consider the choice of secure wireless communication protocols (for data transmission and firmware updates). Moreover, recursive testing should be performed in experimental scenarios that are representative of the various dimensions for a healthy life indoors. The primary goal with this testing is to reach an ideal equilibrium point between IAQ, energy efficiency and thermal comfort.

## 3    Traditional IoT Architectures for Radon Management

This section analyzes traditional IoT technologies for radon gas management to perceive the convergence of the various scientific studies towards the compatibility of their architectures and among the involved different systems, in a medium and long-term vision. Figure 1 depicts a traditional Cloud Computing based communications architecture for radon gas monitoring, which is divided into two main layers:

- **IoT device layer**: it is composed of the different IoT radon gas monitoring sensors and IoT mitigation actuators like fans. Such sensors and actuators are deployed throughout the building and transmit data to a wireless communications router that is connected to the Internet (in Figure 1, it is assumed that IoT nodes make use of Wi-Fi).
- **Cloud**: it collects data through a backend from the deployed IoT sensors/actuators, and processes and stores them in a Cloud database. Such data are then used for making decisions on the mitigation measures and are shown to remote users through an interface like a web-based frontend.

The focus of this work is on prospective analysis, where to anticipate is to act. It is based on uncertainties and on the anticipation of logical scenarios and future events, considering the existence of possible ruptures that can help to dispel doubts regarding the way forward. It is not intended to evaluate quantitative and qualitative factors, but rather to study the risks from a holistic perspective. The prospective strategy is, perhaps, the most important element of this analysis since it allows for taking strategic decisions as an anticipated response to future events. In this way, the prospective analysis will be divided into 2 types of uncertainties:

- **Structural uncertainties**: Based on the cause-effect principle.
- **Unpredictable uncertainties**: Based on the anticipation of sequential chains of events that can turn into disruptive future events. The narrative of future events is extremely difficult to conceive, and it starts with the origin of the idea, always based on scientific credibility, coherence of facts, probability of occurrence, pertinence, and transparency.

### 3.1    Structural uncertainties

Cloud-based communications architectures are currently very popular for deploying IoT applications, but they have a number of drawbacks:

- **Saturation**: As the number of IoT communications grows, the Cloud can become saturated and its performance may decrease dramatically and even stop working at certain moments if mitigation measures are not taken.
- **Scalability**: To address IoT device growth, a Cloud should be scaled, which is not always straightforward, since it requires adding computer servers to already crowded physical spaces and rack servers that cannot be uploaded easily.

- **Internet connectivity dependence**: Internet connectivity is essential to connect IoT radon gas monitoring nodes to the cloud. Therefore, if no Internet connectivity is available (e.g., due to the lack of Wi-Fi coverage in a building), IoT nodes cannot communicate with the Cloud.
- **Single point of failure**: If the Cloud stops working (e.g., due to an internal malfunction, a power outage, a cyberattack, or due to being maintained), the whole IoT system cannot be accessed.

## 3.2    Unpredictable uncertainties

As exploration scenarios that depart from present situations to the future, those related to Cloud-based communications architectures stand out, but it should be noted that they have the following disadvantages:

- **Data privacy and management**: Cloud servers are usually managed by third parties, so data leaks can occur. In addition, many Cloud servers are deployed in foreign countries, whose laws may not guarantee data privacy.
- **Deployment and maintenance cost**: Many Cloud-based solutions are based on third-party services (e.g., Amazon Web Services, Microsoft Azure), which require paying periodic fees. It is possible to deploy private Clouds, but for certain applications, such a deployment and its periodic maintenance may be too expensive.

It is worth considering a fact that is evident: Information and Communication Technology (ICT) industries, building owners, and governments are embracing the concept of smart buildings as a new standard in the world of commercial construction these days. Thus, the future integration of IoT solutions for indoor radon management with Building Automation and Control Systems (BACS) will be feasible. This anticipates the practical implementation of the revision operated by European Directive 2018/844, where the European Union reaffirms its commitment to the development of a sustainable, safe and decarbonized energy system by 2050.

Directive 2018/844 aims to ensure that measures to improve the energy performance of buildings do not focus only on the building envelope but also include all relevant technical elements and systems of a building, which are aimed at reducing energy needs for heating, cooling, lighting, and ventilation. These measures are focused on improving the IAQ and the levels of comfort and well-being of the occupants of the building. In practical terms, it is assumed that this legal framework will be combined with the growing adoption of IP-connected IoT devices to automate unique tasks in smart buildings, which will certainly enhance IAQ-related aspects in future building automation systems, including radon gas mitigation.

Currently, traditional BACS encompasses fully self-contained and largely automated buildings, managed in an integrated manner with the support of Supervisory Control and Data Acquisition (SCADA) systems or specialized industry standards such as LonWorks, BACnet, or KNX. If the paradigm shifts in today's building automation systems to include a permanent interconnection with the building Local Area Network (LAN) or even with the Internet, the cost will be a wider exposure to cyberattacks, which can start inside the building or be initiated anywhere on the  Internet.

Therefore, several technical challenges are foreseen for the main players in this area, such as:

- **IP network interconnection with building automation protocols**: Current BACS include a variety of choices for installation, either wired, wireless, or both. There are several popular connectivity protocols and communication technologies such as GSM, Bluetooth, Wi-Fi, ZigBee, Z-Wave, and KNX. Less popular alternatives include EnOcean, Insteon, LoRaWAN, SigFox, 6LowPAN, Thread and DASH7 [16]. The variety of choices can confuse stakeholders and difficult the right choice, as each of these technologies has some advantages and disadvantages with various technology maturity levels. Yet, it is possible that two or more of these technologies can co-exist in the same building. So, the big challenge will be to converge towards an architecture of compatibility of different building automation protocols with IoT devices, where all protocols can communicate through a common gateway.
- **Smart buildings require security**: Communication networks and transmitted/received data must be completely secure, to prevent unauthorized access and sophisticated cyberattacks that could compromise or disable operations and functionality of smart  buildings.
- **Smart buildings require a low-latency communication infrastructure** (like, for instance, the one based on 5G technologies) to enable the simultaneous interconnection of numerous IoT devices and reconcile the combined computation of sensory technologies, analytical data, Machine Learning (ML), Artificial Intelligence (AI) and Machine-to-Machine (M2M) communication.
- **High-volume data processing**: If the amount of collected and processed data for a given building is very large, it may be necessary to resort to a Big Data system to optimize the computation, which will necessarily increase the cost of the solution in the medium and long term.
- **Standards and certifications**: The establishment of technological standards and certifications must increase feasibility, interoperability, security, privacy, data confidentiality (no one should have access to data without proper authorization), integrity (ensuring that data will not be unduly modified), availability, scalability, and performance to IoT applications, among others.
- **Quality of Service**: It is necessary to define and follow the criteria that guarantee rapid recovery in cases of failures or attacks, protection of communications with redundancy, and rigorous audit protocols and processes.

Considering all these aspects is far from being a trivial task. In addition to the technological challenges, it is essential that all points are treated and shared transparently among the various actors and take into account the global conventions and legislation of each  country.

# 4    Novel Approaches and Future Directions

## 4.1    Edge Computing Architectures

Due to the aforementioned restrictions and risks, new architectures have been recently proposed. Such new architectures are focused on offloading the cloud from the tasks that can be performed in a distributed manner at the network edge. This type of approach is generally called Edge Computing and allows many of the tasks to be carried out on devices close to the end nodes that embed radon gas monitoring sensors.

There are different ways of implementing Edge Computing architectures. When the architecture makes use of devices with significantly high computational power, the devices on the network edge (usually high-performance computers) are called cloudlets [17]. Thus, cloudlets are located close to the end nodes and can process the information received from them [18, 19]. In contrast, when the devices on the network edge have low computing power (e.g., when using Raspberry Pis or other types of Single-Board Computers (SBCs)), the paradigm is called Fog Computing, which was proposed by Cisco in 2012 [20].

Figure 2 shows an example of Edge Computing based architecture for radon gas monitoring through IoT nodes. Three different layers can be distinguished:

- **IoT Device Layer**: It works as in a Cloud-based architecture, but its data are forwarded to the Edge Computing Layer.
- **Edge Computing Layer**: It collects, processes, and stores the information from the IoT Device Layer. In Figure 2, heavy data processing is carried out employing a Cloudlet, while low computing workloads can be handled by Fog Computing gateways. Such gateways can respond fast to the deployed IoT devices and act rapidly on the actuators. Moreover, several gateways can collaborate to perform complex computational tasks.
- **Cloud**: It carries out the tasks that cannot be performed by the Edge Computing Layer and also provides access to remote users through a frontend.

This kind of architecture has contributed to the development of modern IoT systems, but they also suffer from certain drawbacks. Foremost, the deployment of Cloudlets and other Edge Layer gateways in different locations incur additional equipment costs. Also, their maintenance can be complex and expensive. In addition, Edge Layer devices are usually less robust than the cloud and, since they are installed in less controlled environments than a big datacenter, they are exposed to external factors such as changes in temperature or weather conditions that negatively affect their availability. This sort of impact can be a major problem for the performance of the overall system since the Edge Layer is a single point of failure for the area that is being managed. This can lead to connection losses that in IAQ management systems can result in the potential loss of relevant data for an entire area.

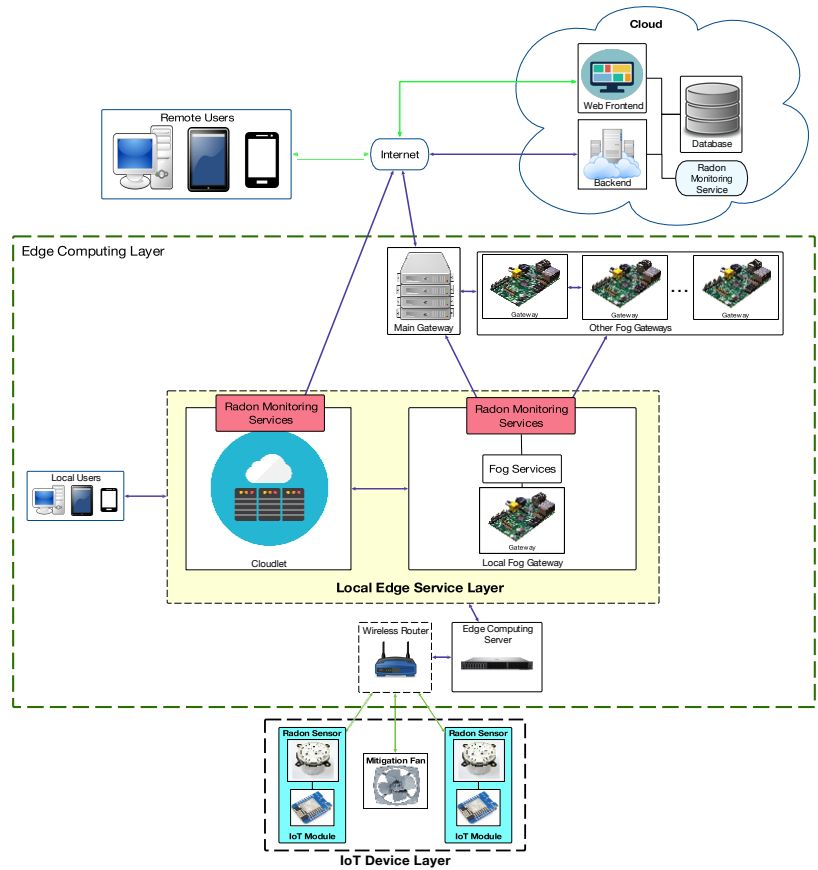

Fig. 2: Example of Edge-based computing and communications architecture for indoor radon management.

## 4.2   Mist Computing and Decentralized Architectures

Another type of architecture relies on offloading network computational capacity directly to the IoT end nodes. Such a computation is usually performed by the microcontrollers or System-on-Chips (SoCs) that collect the information from the deployed sensors [21, 22]. This type of architecture is commonly termed as Mist Computing [23] and allows complex tasks to be performed without the need for delegating them to higher layers thanks to the ability of the mist devices to communicate and coordinate with each other.

Mist Computing systems, thanks to their capacity for not depending  on higher layers, reduce significantly the amount of hardware required to deploy the architecture, which often implies lower costs and less energy consumption. In addition, by  reducing the number of nodes that the requests have to go through to reach the end devices, latency time is also reduced.

Since in this type of architecture, nodes have greater autonomy for data processing and decision-making, the system can continue to operate even if some nodes are out of service. In addition, the opportunity of communicating from node to node makes it possible for communications to occur through the shortest paths or through alternative paths in case of failures.

Figure 3 illustrates a Mist-based computing architecture for indoor radon management that uses IoT nodes with limited computing capabilities that can communicate with each other. As it can be observed in Figure 3, each IoT device has networking capabilities that allow them to exchange information with other IoT devices and to send and receive data directly to and from the cloud through a communications gateway.

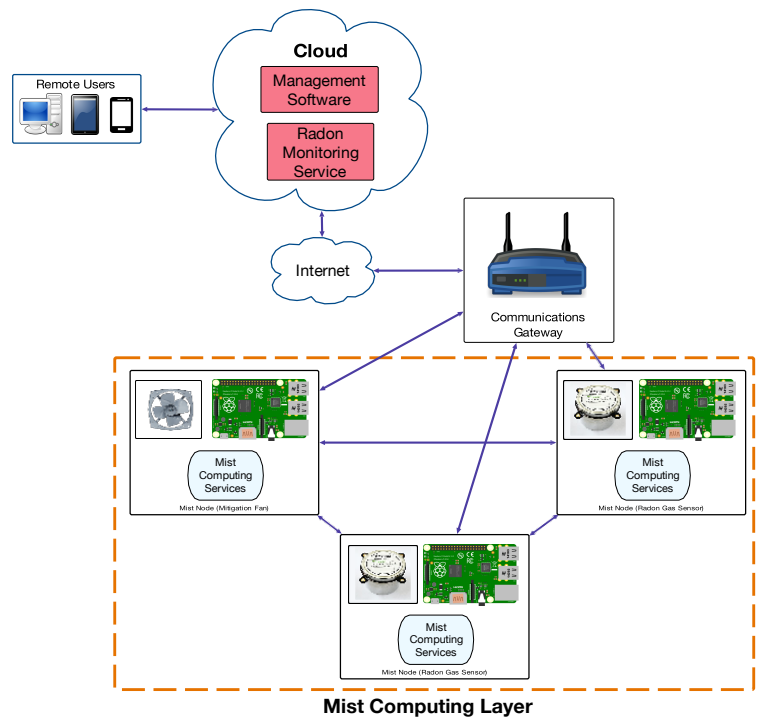

Fig. 3: Example of Mist-based computing and communications architecture for Radon gas monitoring.

Mist-based architectures can be considered decentralized, as most of the computational load is distributed across different devices and end nodes have great autonomy to perform complex tasks. However, most mist architecture still has a centralized element that is often a service running in the cloud that performs co-ordination tasks and provides access to external applications or user interfaces. To avoid this limitation, recently novel completely decentralized architectures are emerging. In these architectures all nodes perform their role in accordance with a decentralized protocol. Such architectures seek to eliminate as much as possible the dependency on external services so that no network node is absolutely necessary and all of them cooperate to achieve the same purpose making use of specialized protocols to achieve decentralized networking, computing, and/or storage [24]. This allows decentralized networks to have a very high availability without significantly increasing the material cost.

Typically, the different types of edge architectures are not mutually exclusive, and, depending on the specific application, different types can be used together to obtain the required features [25]. Thus, a system can make use of both Fog Computing devices and Cloudlets, which handle the most complex computational processes, and a Mist Computing layer that is able to manage the collected data autonomously.

### 4.3   Desired Features for Next-Generation Radon Gas Monitoring Systems

Regardless of the type of architecture, IAQ management systems have a number of specific characteristics:

- End devices are usually spatially scattered through an extensive area (e.g., a building, an industrial complex, a campus).
- Some devices may be located in hard-to-reach or restricted access areas (e.g., machine rooms, warehouses, datacenters).
- Some nodes may be located in environments where access without protective equipment is discouraged.
- The nodes may be connected to the Internet through a third-party network that may suffer interruptions due to failures or to maintenance works.

Due to these specific characteristics of IAQ monitoring systems, there are a number of desirable functionalities to mitigate the problems associated with the environments where the nodes are deployed. These functionalities ease their maintenance, their audits, and the continuous improvement of the IoT network:

- **Remote monitoring and self-diagnosis systems**: This feature makes it possible to detect problems and know the status of the network remotely without needing to physically access the different nodes of the network.
- **Local data storage system and delayed information forwarding**: This feature enables the data obtained from the sensors can be stored at the end node itself to prevent losses of information in the event of network interruptions.
- **Remote update system**: This system allows for performing firmware updates of the nodes through the network in order to improve and to keep the system up to date without the need for physically visiting or replacing all the deployed nodes, which would have a high cost.
- **System reaction to planned events**: The deployed radon gas monitoring nodes must be programmed to react to certain planned events and to be able to make simple decisions without relying on third parties to mitigate a possible network outage.

Finally, it is worth mentioning that there are a number of research lines on radon gas and IAQ management systems that are currently gaining interest:

- **Full decentralization**: Fully decentralized IoT systems are being proposed to provide decentralized data sharing systems and consensus protocols for the network [26, 27].
- **Energy consumption optimization**: Each node in an IoT system usually consumes an almost negligible amount of power, but the deployment of thousands of IoT nodes globally has a considerable impact on energy consumption and resource management. Thus, energy awareness is a topic that is gaining more and more interest, as well as the creation of more efficient and sustainable IoT networks [28, 29].
- **Energy harvesting systems**: Along with efficient use of resources, harvesting energy from alternative sources is a topic that is gaining more and more interest to power devices in remote areas for long periods of time or to facilitate energy supply to low-power devices using sources available in their environment (e.g., piezoelectric, thermal, RF) [30].
- **Digital Twin integrations**: The standardization of data format protocols is enabling the integration of multiple sensors into complex systems that enable advanced data analysis in what is known as digital twins. This term is being widely studied in Industry 4.0 [31], but recently it is also gaining interest in home automation environments.

## 5   Conclusions

This article reviewed the evolution of IoT architectures for indoor radon gas management (which includes both monitoring and mitigation) by providing a holistic approach to the topic. The main characteristics of IoT technologies for indoor radon management were described, and the most relevant subsystems were detailed, as well as their main communications architectures. In addition, the most relevant academic works related to indoor radon management solutions have been analyzed, showing the potential of such IoT-enabled systems. Furthermore, after describing traditional IoT architectures, novel approaches like Edge, Mist, and decentralized computing paradigms, have also been analyzed in order to emphasize their potential for creating next-generation indoor radon management solutions. Finally, the main open challenges and future directions for the development of such solutions have been enumerated. As a result, this article provided useful guidelines for the IoT designers and developers of future indoor radon management systems.

## Acknowledgments

This work is a result of the project TECH—Technology, Environment, Creativity and Health, Norte-01-0145-FEDER-000043, supported by Norte Portugal Regional Operational Program (NORTE 2020), under the PORTUGAL 2020 Partnership Agreement, through the European Regional Development Fund (ERDF).

The authors would like to thank CITIC for its support for the research stay that led to this article. CITIC, as Research Center accredited by Galician University System, is funded by "Consellería de Cultura, Educación e Universidades from Xunta de Galicia", supported in an 80% through ERDF Funds, ERDF Operational Programme Galicia 2014-2020, and the remaining 20% by "Secretarı́a Xeral de Universidades" (Grant ED431G 2019/01). This work has also been funded by the Xunta de Galicia (by grant ED431C 2020/15), the Agencia Estatal de Investigación of Spain, MCIN/AEI/10.13039/501100011033 (by grant PID2020-118857RA-I00 (ORBALLO)) and ERDF funds of the EU (FEDER Galicia 2014–2020 & AEI/FEDER Programs, UE).